\documentclass[10pt, conference]{IEEEtran}
\IEEEoverridecommandlockouts

\usepackage{lettrine}
\usepackage{amsfonts}
\usepackage[caption=false,font=normalsize,labelfont=sf,textfont=sf]{subfig}
\usepackage{indentfirst}
\usepackage{makecell}
\usepackage{comment}

\usepackage{etoolbox}
\usepackage{array}
\usepackage{algorithm}
\usepackage{algpseudocode}
\usepackage{float}
\usepackage{tabularray}
\usepackage[numbers]{natbib}
\usepackage{multicol}
\usepackage{multirow}
\usepackage[letterpaper,top=2cm,bottom=2cm,left=3cm,right=3cm,marginparwidth=1.75cm]{geometry}
\usepackage{graphicx}

\usepackage{xcolor}
\usepackage{amsmath}
\usepackage[colorlinks=true, allcolors=blue]{hyperref}
\usepackage{arydshln}
\usepackage{float}

\newcommand{\vars}{\texttt}
\newcommand{\func}{\textrm}
\newcolumntype{x}[1]{>{\centering\arraybackslash\hspace{0pt}}p{#1}}

\begin{document}

\title{Optimized Deep Learning Models for Malware Detection under Concept Drift}

\author{
\IEEEauthorblockN{
William MAILLET and 
Benjamin MARAIS\IEEEauthorrefmark{1}}
\vspace{0.1cm}

\IEEEauthorblockA{Orange Innovation, Rennes, France}
\IEEEauthorblockA{\footnotesize{mails : [william.maillet, benjamin.marais]@orange.com}}
\IEEEauthorblockA{\footnotesize{\IEEEauthorrefmark{1}corresponding author}}}

\maketitle

\begin{abstract}
Despite the promising results of machine learning models in malicious files detection, they face the problem of concept drift due to their constant evolution. This leads to declining performance over time, as the data distribution of the new files differs from the training one, requiring frequent model update. In this work, we propose a model-agnostic protocol to improve a baseline neural network against drift. We show the importance of feature reduction and training with the most recent validation set possible, and propose a loss function named  Drift-Resilient Binary Cross-Entropy, an improvement to the classical Binary Cross-Entropy more effective against drift. We train our model on the EMBER dataset, published in2018, and evaluate it on a dataset of recent malicious files, collected between 2020 and 2023. Our improved model shows promising results, detecting 15.2\% more malware than a baseline model.
\end{abstract}

\vspace{0.2cm}
\begin{IEEEkeywords}
 Artificial intelligence, Deep learning, Concept drift, Neural Network, Malware detection
\end{IEEEkeywords}

\section{Introduction}

Traditional malware detection methods rely on signatures, heuristics and behaviors \cite{Aboaoja2022,Bazrafshan2013}. However, these solutions are not suitable in the long term due to the significant number of malware present in the cyberspace, and creating new rules for detection becomes an impractical and unscalable approach. As an alternative, machine learning models have demonstrated great success in various tasks, such as classification, computer vision, and anomaly detection, making them promising solutions for the future of malicious software detection. In particular, neural networks and LightGBM \cite{Ke2017} have shown particularly encouraging results \cite{Akhtar2022,Gao2022,Alkasassbeh2020}. This machine learning models can use static characteristics extracted from malicious files, such as imports, strings, and headers information, or dynamic characteristics, as network activity or registry modifications, collected during files execution. While these models perform well, they face the challenge of constant malware evolution. In fact, in 2022, SonicWall reported discovering over $465,000$ new malware variants \cite{SonicWall}. These new files could cause a phenomenon called ``concept drift" \cite{Lu2019} that occurs when the distribution of data changes over time. When a model suffers from concept drift, this can have an impact on its efficiency and cause a drop in performance. Concept drift poses a real problem as it necessitates frequent model retraining, resulting in time and resource losses. Moreover, in the field of malicious files detection, this can lead to security breaches and vulnerabilities.

In this paper, we show that it is possible to improve a baseline model in order to reduce performance degradation due to drift and we propose three model-agnostic contributions. We first suggest using a validation set composed of the most recent malware and benign samples rather than random ones. We propose an enhancement to the classical Binary Cross-Entropy (BCE) loss function, strengthening it against drift using spectral decoupling and penalization of both false negatives and false positives. Lastly, we show that using a subset of features selected through a feature reduction method yields better results. To assess our contributions, we will evaluate our model performance, on two datasets, BODMAS \cite{BODMAS2021} and a hand-collected recent malware dataset, after training it on EMBER \cite{EMBER2018}.

This paper is organized as follows: Section \ref{sec:relevant_works} defines the drift and reviews related research works, Section \ref{sec:methodology} depicts our complete methodology to train and evaluate our models and Section \ref{sec:exp_res} presents the experiments and results. Finally, in Section \ref{sec:conclusion}, we discuss about our work and possible improvements.

\section{Definitions and Related Works}
\label{sec:relevant_works}

In this section, we first define the notion of drift before exploring how it occurs in the malware context. We then summarize state-of-the-art drift detection methods and solutions in the malware detection field.

\subsection{Drift Definition} 

\begin{figure}[!ht]
    \centering
    \includegraphics[scale=0.4]{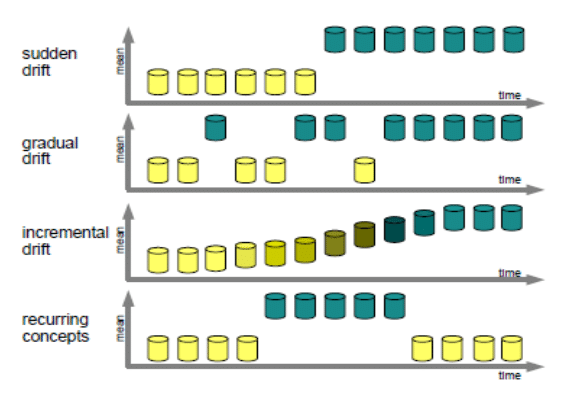}
    \caption[Possible drifts]{Different ways the drift occurs \cite{Lemaire2015}}
    \label{fig:drifttypes}
\end{figure}

A machine learning model is said to be subject to drift when its performance decreases over time. This can happen because the input data distribution is shifting, or even because the relationship between inputs and outputs changes over time. There are multiple ways in which the drift may occur: changes can happen suddenly, gradually, and incrementally but also be recurrent over time as shown in Figure \ref{fig:drifttypes}.

Extensive research has been conducted on the phenomenon of drifts, defining multiple types of drifts and data distribution shifts. In their review, Lu et al. \cite{Lu2019} provide an in-depth exploration of the drift landscape, offering appropriate definitions. In this paper, we focus on investigating and addressing the performance loss of a machine learning model over time and consequently give a specific definition of drift for this purpose.

Let $\theta^*$ be the optimized parameters of a predictive model $f_{\theta^*}$, trained on a set $S_t = (X_t, y_t)$ of observations collected on a period before time $t$. The error of this model is measured by an error function $E$, and we define $\epsilon \in \mathbb{R}$ a fixed error threshold. The model suffers a loss of performance known as drift if their existes $t_0 \in [t, +\infty[$ such as:

\begin{equation}
\begin{split}
E(f_{\theta^*}(X_{[t, t_0[}); y_{[t, t_0[}) \leq \epsilon  \\ 
\text{and} \ \ \ E(f_{\theta^*}(X_{[t_0, +\infty[}); y_{[t_0, +\infty[}) \geq \epsilon
\end{split}
\end{equation}

This definition implies that the drift must be measured using an error metric, which is not always possible.  Indeed, for a company trying to classify never-before-seen samples, it is often impossible to evaluate model accuracy without human intervention to study and label new samples. Multiple drift detection techniques have been studied \cite{GoncalvesJr2014,Gemaque2020}, but, in this work, we will only focus on methods that reduce and prevent it.

\subsection{Drift in Malware Context} 

In the context of malware detection, we can distinguish several sources of drift, which affect machine learning models in different ways. The most common one is the appearance of a new type or a new family of malware \cite{Limin2021}. This type of drift could be classified as a sudden drift as new data samples appear in the input distribution. Attackers often work as groups or organizations and are likely to follow trends depending on which malicious files are the most used and efficient at a given time. These changes in the prevalence of malware families \cite{Kan2021} could be categorized as both gradual and recurrent drift, as the distribution of malicious software will follow trends. It is also worth mentioning the possible updates of API and libraries \cite{Zhang2020}. Indeed, the API calls used by benign and malicious files can change if a major update takes place. This event could affect both malicious and benign files in the same way. Interestingly, sources of drift in the malware context affect only the input features distribution. This type of data distribution shift where the distribution of inputs change over time is commonly referred to as covariate shift, or virtual drift \cite{Huyen2022,Lu2019}.

\subsection{Solutions to Handle Drift}
\label{sec:solutions_against_drift}

In order to address drift issues, the initial step frequently involves detecting or highlighting its existence. Detecting drift poses a significant challenge in many existing studies, and several methods for detecting drift in this purpose have been developed, depending on their suitability for the specific drift at hand. Several methods use distance-based metrics over the input features \cite{Limin2021,Singhal2020} to detect drifting samples. Another common way to detect drift is by monitoring the performance of the classifier over time \cite{Jordaney2017,Barbero2022,Hu2017}.

Many studies focus on the learning process of models in order to prevent drift. Online-learning solutions are very popular \cite{Narayanan2016,Huynh2017,Xu2019} as they yield good results against drift and are easily integrated in working environments. Contrastive learning was also proven to be useful \cite{Limin2021,Chen2023} as it teaches the model how to identify the differences or similarities between two samples. Another approach being studied is the use of artificial samples, such as Tumpach et al. \cite{Tumpach2019} that proposed a learning framework using variational autoencoders (VAE) trained to regularly generate features for the model helping it to be more robust.

Among the learning methods with a positive impact on model performance, the loss function penalization called ``spectral decoupling" proposed by Pezeshki et al. \cite{Pezeshki2021} has recently been shown to help neural networks to deal with concept drift \cite{Chalkidis2022} and data distributions shifts \cite{Pohjonen2022}. The BCE loss with spectral decoupling is defined as:

\begin{equation}
    \label{eq:1}
    \mathcal{L}_{SD}(p, y) = \mathcal{L}_{BCE}(p, y) + \frac{\lambda}{2} * ||z||^2_{2}
\end{equation}
where $y \in \{0, 1\}$ is the ground-truth class of the sample, corresponding to ``malware" if $y=1$, and ``benign" otherwise. The value $z \in \mathbb{R}$ denotes the raw outputs of the model (logits), and  $p \in [0,1]$ is the model's output probability prediction (i.e. $P(y=1)$) defined as $p = \sigma(z)$ with $\sigma$ the sigmoid function.

\section{Methodology}

\label{sec:methodology}
In this section, we present our methodology to improve a baseline neural network model to handle the drift problem. We first describe the datasets used along with our training protocol, and we present our improvements and describe our models.

\subsection{Data and Training Approach}

\begin{figure}[H]
    \centering 
    \includegraphics[scale=0.25]{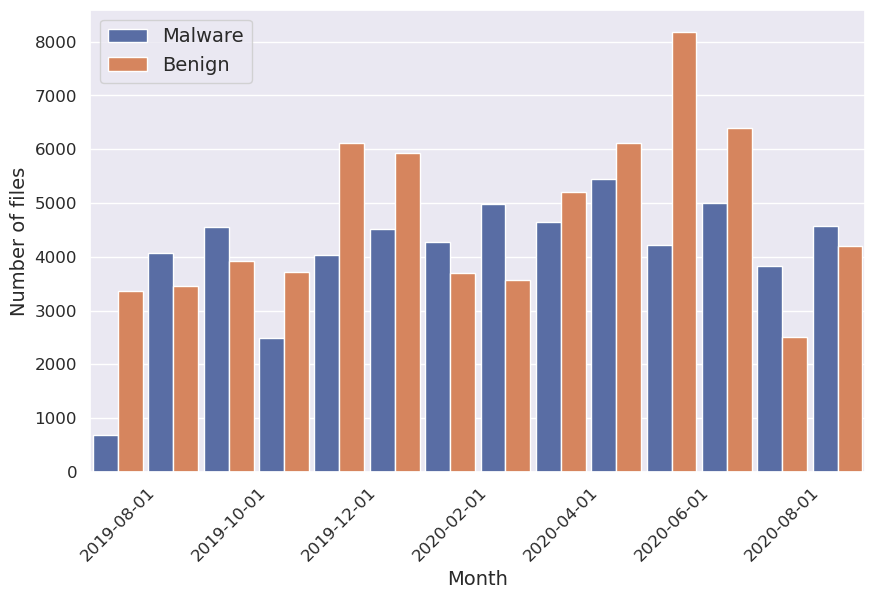}
    \caption[BODMAS monthly malicious and benign files distribution]{BODMAS monthly malicious and benign files distribution}
    \label{fig:BODMAS}
\end{figure}

To carry out our experiments, we used two public datasets and a hand-collected one. The first public dataset we used is EMBER \cite{EMBER2018} which is a well-known dataset in the field of malware detection, that provides extracted feature vectors for benign and malicious files, collected between 2017 and 2018. 
Secondly, we used the BODMAS dataset \cite{BODMAS2021} which is more recent as the files are gathered between 2019 and 2020, and authors also use the feature extraction method provided by EMBER. As we seek to visualize the performance evolution over time, the set is divided by month, resulting in 14 consecutive subsets from August 2019 to July 2020, as shown in Figure \ref{fig:BODMAS}.

\begin{figure}[H]
    \centering 
    \includegraphics[scale=0.25]{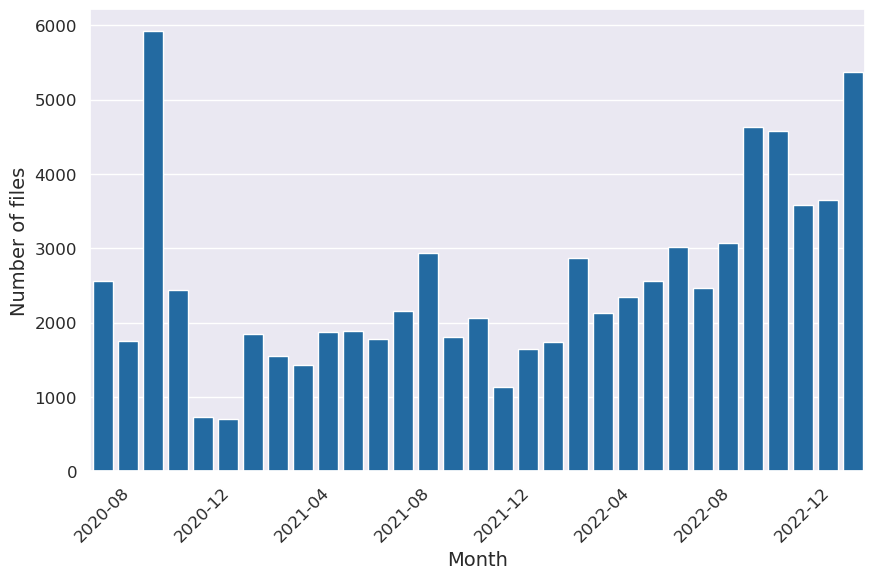}
    \caption[Monthly sample distribution in the MalwareBazaar dataset]{Monthly sample distribution in the MalwareBazaar dataset}
    \label{fig:mwb}
\end{figure}

To build the third dataset, we hand-collected malware submitted on the MalwareBazaar website \cite{MalwareBazaar} between 2020 and 2023. We transform these raw files into vectors using the EMBER feature extractor. We also divided this dataset in multiple consecutive subsets for each month. Figure \ref{fig:mwb} shows the distribution of the MalwareBazaar dataset over time.  

The EMBER and BODMAS datasets are made up of instances collected in a similar context, whereas the MalwareBazaar instances, in addition to being chronologically more distant, are collected on a collaborative, unmoderated database. Moreover, MalwareBazaar contains no benign files, making it a dataset that is semantically quite distant from the other two.

\begin{figure*}[!ht]
    \centering 
    \includegraphics[scale=0.7]{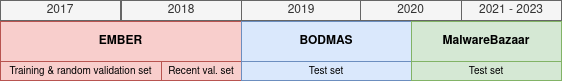}
    \caption[Illustration of the subsets]{Illustration of the chronological order for each subset}
    \label{fig:subset}
\end{figure*}

The purpose of our works is to minimize the impact of the drift on malware detection models over time. We seek to place ourselves in a condition close to the reality of a company, and we therefore propose a framework to train our malware detection models in this setting. The training set we used is a subset of EMBER containing $700,000$ random instances from 2017 and 2018. To evaluate the performance of our model, we use the BODMAS dataset (130,000 samples) and our MalwareBazaar dataset ($78,000$ malware samples) as test sets. Our intuition is that by training a model on an older dataset (EMBER) and evaluating it on more recent data (BODMAS and MalwareBazaar datasets), we expect to highlight a decline in model performance due to data drift over time. Moreover, during the learning process, we use two different validation set composed of $120,000$ instances. The first set consist of randomly selected EMBER samples, and the second set is composed of the most recent EMBER instances. By using a more recent subset for the validation process, we expect to avoid overfitting on old data, improve model generalization and reduce drift effect. Figure \ref{fig:subset} summarizes the chronological order and use of each dataset.

\subsection{Feature Selection and Reduction}
\label{sec:perm_feat_imp}

Feature reduction has been shown to help models generalization capacities \cite{Kumar2014,Venkatesh} and decrease computational costs, which can be a limiting factor in companies. Moreover, as described in the discussion section of the EMBER paper, the extracted features mix all sorts of information that might not be relevant to a malware detection task.

The vectors provided by EMBER are composed of $2,381$ static features including statistical variables as bytes distribution and entropy, or files contents as headers information and import functions. We chose to run all our experiments without the timestamp variables because it is known to be unreliable and easy to alter for an attacker \cite{PEFileTimeStamp}.

For our experiments, we then chose to apply the Permutation Feature Importance (PFI) algorithm to our models to reduce the number of features extracted using the EMBER extractor. PFI has been introduced by Breiman \cite{Breiman2001} and refined by Fisher et al. \cite{Fisher2019}. This algorithm computes the difference between the model's performance on the regular dataset and on a dataset where a feature is randomly permuted along the samples. If the performance deteriorates, the feature should be kept as the model relies on it for its predictions. Randomly permuting a feature is a mean of breaking the relationship between the feature and the output, while keeping inputs from the same distribution \cite[Ch.\ 8.5]{Molnar2022}. Our intuition is that feature reduction will help in identifying characteristics that are relevant over time, and removing those that are significant in old datasets, but perhaps less for detecting recent malicious files. The PFI algorithm is described in Algorithm \ref{algo:pfi}.

\begin{algorithm}[!ht]
\caption{Permutation Feature Importance Algorithm}\label{algo:pfi}
    \begin{algorithmic}
        \Function{PFI}{$\vars{M}, \vars{X}, \vars{y}, \vars{E}$}   \Comment{\vars{M} the trained model, \vars{X} the input matrix, \vars{y} the target vector, \vars{E} the desired metric to perform evaluation}
        
        \State $\vars{E}_{base} \gets \vars{E}(\vars{M}(\vars{X}), \vars{y})$
        \State $\vars{L} \gets \vars{[]}$ \Comment{The list of feature indices to keep}
        \For{$\vars{i}\gets 1, \vars{N}_{feat}$} \Comment{$\vars{N}_{feat}$ the number of features}
            \State $\vars{X}_{prm}$ = \func{RandomPermuteFeature}(\vars{X}, \vars{i}) \Comment{Permuting a feature along the samples}
            \State $\vars{E}_{prm} = \vars{E}(\vars{M}(\vars{X}_{prm}), y)$
        \If{$\vars{E}_{base} - \vars{E}_{prm} > 0$}
            \State $\vars{L}.$\func{Insert}($\vars{i}$)
        \EndIf
        \EndFor
        \State \Return $\vars{L}$
        \EndFunction
    \end{algorithmic}
\end{algorithm}

\subsection{Drift Resilient Binary Cross-Entropy}
\label{sec:DRBCE}

We propose a custom loss function based on the BCE loss, combining three major components. The first component is spectral decoupling \cite{Pezeshki2021}, which is a loss regularization term known to improve generalization and robustness against drift (see \ref{sec:solutions_against_drift}). We also introduce false positives (FP) and false negatives (FN) penalization, a common technique that adds penalization terms to logistical losses biasing the model towards learning with less FP or FN predictions. This technique has been used in the malware detection context by \cite{Gao2022}, yielding improved results on a LightGBM model \cite{Ke2017}. Our third component is class weighting. It is used to adapt the loss for unbalanced datasets during training.

This custom loss function, denoted as ``Drift Resilient Binary Cross-Entropy" (DRBCE) loss, is defined as:

\begin{equation}
    \label{eq:drbce}
    \begin{split}
            \mathcal{L}_{DRBCE} = &- \frac{1}{N} \sum_{i=1}^{N}  [ w_1 \cdot P_{FN} \cdot y_i \cdot log(p_i) \\
            &+ w_0 \cdot P_{FP} \cdot (1 - y_i) \cdot log(1 - p_i)] \\
            &+ \frac{1}{N}  \sum_{i=1}^{N} \frac{\lambda}{2} ||z_i||^{2}_{2}
    \end{split}
\end{equation}
where $N \in \mathbb{Z}$ denotes the number of samples, $P_{FN}, P_{FP} \in \mathbb{R}$ are respectively the FN and FP coefficients \cite{Gao2022}, $y_i \in \{0, 1\}$ is the ground-truth class of the sample $i$ associating $y=1$ to malware and $y=0$ to benigns, $z_i \in \mathbb{R}$ is the raw output of the model (logits) for sample $i$, $p_i \in [0,1]$ is the model's probability prediction for class $y=1$, $w_1, w_0 \in [0, 1]$ are respectively the class weights of class $y=1$ and $y=0$ and $\lambda \in \mathbb{R}$ is the spectral decoupling coefficient. The model's probability prediction $p_i$ is defined as $p_i = \sigma(z_i)$ with $\sigma$ the sigmoid function. The class weights are defined as $w_1 = \frac{N_1}{N}, w_0 = \frac{N_0}{N}$, with $N_k$ the number of samples with class label $y=k$.

The values of hyper-parameters $P_{FN}$, $P_{FP}$, and $\lambda$ are chosen during the experimentation phase with hyperparameters tuning.

\subsection{Models}
\label{sec:model}

To carry out our experiments, we define a baseline model to which we apply gradual modifications. We then compare our baseline model with its variants to assess the performance of our contributions. We first describe our baseline model and then list the experiments and modifications that we are going to carry out.

\begin{figure}[!ht]
    \centering         
    \includegraphics[scale=0.6]{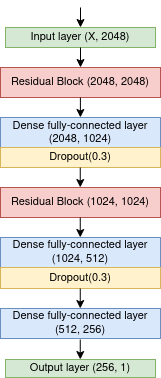}
    \caption[Architecture of the baseline model]{Architecture of the baseline model}
    \label{fig:model}
\end{figure}

Our baseline model is a Residual Neural Network (RNN). It is built from two residual blocks and dense fully-connected layers, using the ReLU activation function and dropout, as described in Figure \ref{fig:model}. We chose to use a classical architecture because our contributions are model agnostic, meaning they should work for classical neural network architecture, as well as other type of algorithm, such as Lightgbm for instance.

The baseline model is trained using the BCE loss function, and the AdamW optimizer with a learning rate of $1 \times 10^{-4}$ and a weight-decay coefficient of $1 \times 10^{-4}$ \cite{Ilya2019}.

During our experiments, we tested three main changes on the baseline model: the validation set, the loss function and the feature reduction. We experimented replacing the random validation dataset with a set  that contains the latest samples from EMBER. Thus, we expect to avoid overfitting on older features. We also tested replacing the BCE loss function with our DRBCE function, aiming to improve the performance of our model. Finally, we also performed PFI on our dataset with the baseline model, and trained our model with reduced features, seeking to enhance the generalization of our model. After applying the PFI algorithm, we adapted the size of the input layer to fit the new number of features.

\subsection{Evaluation}
\label{sec:eval}
We evaluate our models using well-known metrics as accuracy (ACC), F1-Score (F1), False Positive Rate (FPR), and False Negative Rate (FNR). In addition to accuracy score, measuring the number of files correctly classified, we also use F1-Score, FNR and FPR to get more insights on our models performance. The F1-Score is interesting in the case of unbalanced datasets as it combines precision and recall into a single metric. FPR and FNR are necessary metrics in the malware detection context: depending on the application, one might prefer detecting more malicious files and accepting getting more false positive predictions rather than detecting fewer malware and accepting getting more false negative predictions. 
In the context of a security operations center, the company may seek to decrease the FPR since excessive handling of events can result in alert fatigue \cite{Tao2021}. On the other hand, within the domain of an antivirus solution, the company usually focuses on enhancing the detection of malicious files, thus prioritizing a low FNR. In this study, we focuse on minimizing the FNR to achieve the highest possible detection of malware instances.

\section{Experiments and Results}
\label{sec:exp_res}

\subsection{DRBCE Loss and hyper-parameters tuning}

In this section, we describe our experiments and results with the DRBCE loss, defined in Section \ref{sec:DRBCE} as: 
\begin{equation}
\centering
    \label{eq:drbce_bis}
\begin{split}
\mathcal{L}_{DRBCE} = &- \frac{1}{N} \sum_{i=1}^{N}  [ w_1 \cdot P_{FN} \cdot y_i \cdot log(p_i) \\ 
&+ w_0 \cdot P_{FP} \cdot (1 - y_i) \cdot log(1 - p_i)] \\ 
&+ \frac{1}{N}  \sum_{i=1}^{N} \frac{\lambda}{2} ||z_i||^{2}_{2}
\end{split}
\end{equation}

To get the best performance out of this loss function, we seek to fine-tune the three hyper-parameters $P_{FP}, P_{FN} \in \mathbb{R}$ respectively the false positives and false negatives penalization coefficients and $\lambda \in \mathbb{R}$ the spectral decoupling coefficient.

\subsubsection*{Spectral Decoupling Coefficient Tuning}

We first seek to find the best hyper-parameter $\lambda$ value for the spectral decoupling of our DRBCE loss function to help the model get the best generalization possible. We test several values of $\lambda$ for a fixed neutral penalization pair $(P_{FN}=1, P_{FP}=1)$ and observe the F1-Score, FNR and FPR of the model.

\begin{table}[!ht]
    \caption{Models performance for different $\lambda$ coefficient values}
    \def\arraystretch{1.1}
    \centering 
    \begin{tabular}{ |c | c | c | c | c | c |}
    \hline
    \multirow{2}{*}{\textbf{$\lambda$}} & \multicolumn{3}{c|}{\textbf{BODMAS}} & \multicolumn{2}{c|}{\textbf{MalwareBazaar}}\\
    \cline{2-6} 
        & \textbf{ACC} & \textbf{F1} & \textbf{FNR} & \textbf{ACC} & \textbf{F1}\\
    \hline 
        $\lambda = 0.5$ & 0.8806 & 0.8469 & 0.1606 & 0.3191 & 0.4694\\
        $\lambda = 0.1$ & 0.9370 & 0.9274 & 0.0550 & \textbf{0.3715} & \textbf{0.5410} \\
        $\lambda = 0.05$ & \textbf{0.9422} & \textbf{0.9337} & \textbf{0.0468} & 0.3621 & 0.5314 \\
        $\lambda = 0.01$ & 0.9379 & 0.9272 & 0.0708 & 0.3517 & 0.5202\\
        $\lambda = 0.001$ & 0.9255 & 0.9137 & 0.0744 & 0.3642 & 0.5337\\        
    \hline
    \end{tabular}
    \label{tab:lambda}
\end{table}

Although $\lambda = 0.05$ gives the best results on the BODMAS dataset, we would rather use a $\lambda$ value that would give us the best performance on recent malware, while maintaining a fairly acceptable accuracy on the BODMAS dataset. Consequently, we are opting for $\lambda = 0.1$ for the rest of the experiments.

\subsubsection*{Penalization coefficient tuning}

We then compare multiple penalization pairs $(P_{FN}, P_{FP})$, to observe their impact on models performance. Our intuition is that these penalization coefficients help the model integrating the notion of false negatives and false positives instead of focusing only on achieving the best possible accuracy \cite{Gao2022}. Table \ref{tab:penalization} gives ACC, F1, FNR, and FPR for different pairs, with fixed $\lambda = 0.1$. 

\begin{table}[!ht]
    \caption{Models performance for different penalizations pairs}
    \def\arraystretch{1.1}
    \centering 
    \begin{tabular}{ |c | c | c | c | c | }
    \hline
    \multirow{2}{*}{\textbf{$(P_{FN}, P_{FP})$}} & \multicolumn{4}{c|}{\textbf{BODMAS}}\\
    \cline{2-5} 
        & \textbf{ACC} & \textbf{F1} & \textbf{FNR} & \textbf{FPR}\\
    \hline 
        $P_{FN} = 1, P_{FP} = 1$ & 0.9370 & 0.9274 & 0.0550 & 0.0690 \\
        $P_{FN} = 1, P_{FP} = 3$ & \textbf{0.9406} & \textbf{0.9294} & 0.0825 & \textbf{0.0465} \\
        $P_{FN} = 3, P_{FP} = 1$ & 0.9227 & 0.9142 & \textbf{0.0342} & 0.1094  \\
        $P_{FN} = 5, P_{FP} = 1$ & 0.9079 & 0.8995 & 0.0348 & 0.1346 \\        
    \hline
    \end{tabular}
    \begin{tabular}{ |c | c | c | c |}

    \multirow{2}{*}{\textbf{$(P_{FN}, P_{FP})$}} & \multicolumn{3}{c|}{\textbf{MalwareBazaar}}\\
    \cline{2-4} 
        &\textbf{ACC} & \textbf{F1} & \textbf{FNR}\\
    \hline 
        $P_{FN} = 1, P_{FP} = 1$ &  0.3715 & 0.5410 & 0.6285\\
        $P_{FN} = 1, P_{FP} = 3$ &  0.3043 & 0.4659 & 0.6765\\
        $P_{FN} = 3, P_{FP} = 1$ & 0.4605 & 0.6300 & 0.5395\\
        $P_{FN} = 5, P_{FP} = 1$ & \textbf{0.4877} & \textbf{0.6957} & \textbf{0.5123}\\        
    \hline
    \end{tabular}

    \label{tab:penalization}
\end{table}

\begin{figure*}[htb]
    \centering 
    \subfloat[FNR]{
        \includegraphics[scale=0.25]{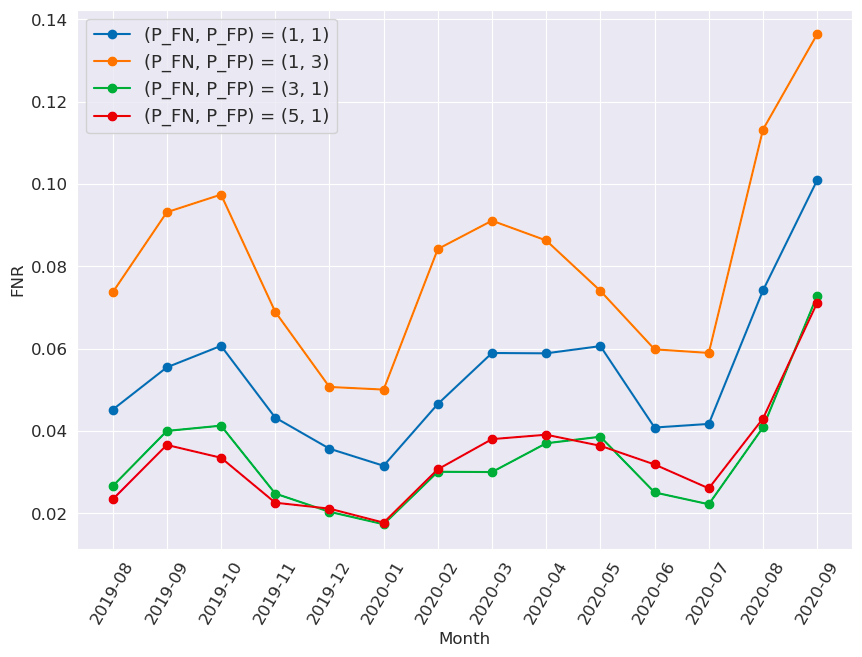}\label{fig:penalization:b}}
    \subfloat[FPR]{
        \includegraphics[scale=0.25]{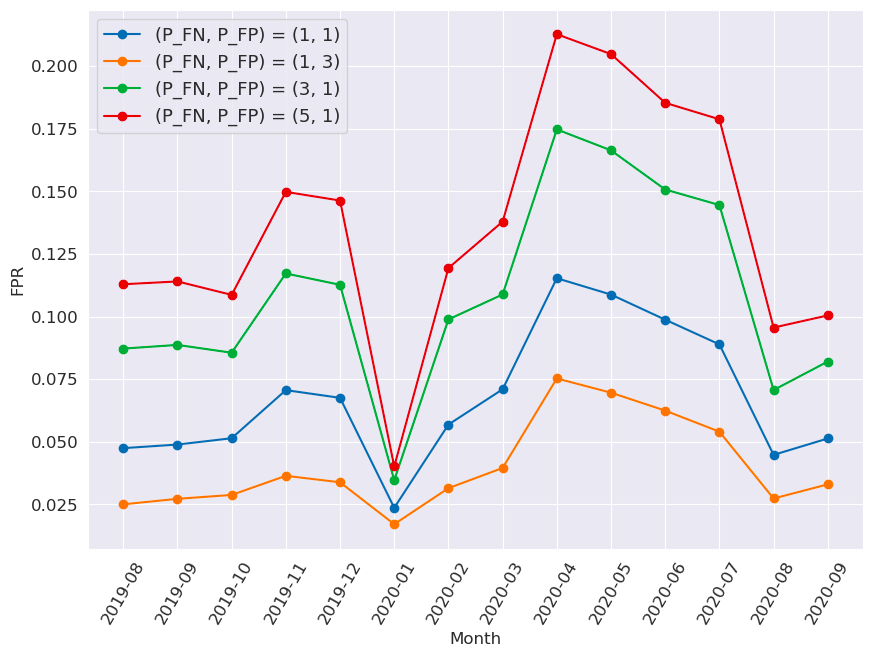}\label{fig:penalization:c}}
    \caption[Models performance comparison for different penalization pairs]{Metrics comparison for different penalization pairs}
    \label{fig:penalization}
\end{figure*}

When observing the models with more penalization on either FP or FN, we can see that increasing the penalization on FN increases the FPR, and conversely as shown in Table \ref{tab:penalization} and Figure \ref{fig:penalization}. This relationship between FP and FN penalization also impacts the global classification performance of the model, as the best performing pair $(P_{FN}, P_{FP})$ on the most recent dataset is $(5, 1)$, meaning that penalizing five times more FN predictions results in a much better handling of drift (only a 2.9\% accuracy decrease on BODMAS for nearly a 12\% increase on the MalwareBazaar dataset compared to $(1, 1)$ penalization). As our aim is to have the best long-term performance, we have chosen to use $(5, 1)$ as penalization pair even though the values $(3, 1)$ yields very good results.

\subsubsection*{Final hyper-parameters and comparison}

After our experiments, we chose to set the hyper-parameters of our loss function as $P_{FN}=5$, $P_{FP} = 1$ and $\lambda = 0.1$. As shown in Table \ref{tab:loss_function}, our model performs far better on MalwareBazaar (12.57\% increase in accuracy), and slightly worse on BODMAS (1.6\% accuracy decrease), while reducing the number of false negatives by more than half. These results indicate that the BCE loss function gives better results in the short-term, while our custom loss function improves our model's performance in the long-term, with a slight trade-off in terms of accuracy for samples closest to the training dataset. Our custom loss function allows the users to select penalties based on their detection policy, as described in Section \ref{sec:eval}.

\begin{table}[!ht]
    \caption{Models performance for both BCE and DRCBE losses}
    \def\arraystretch{1.1}
    \centering 
    \begin{tabular}{ |c | c | c | c|}
    \hline
    \multirow{2}{*}{\textbf{\makecell{Loss \\Function}}} & \multicolumn{3}{c|}{\textbf{BODMAS}}\\
    \cline{2-4} 
        & \textbf{ACC} & \textbf{F1} & \textbf{FNR}   \\
    \hline 
        BCE & \textbf{0.9247} & \textbf{0.9127} & 0.0769 \\
        DRBCE & 0.9079 & 0.8995 & \textbf{0.0348} \\
    \hline
    \end{tabular}

        \begin{tabular}{ |c | c | c | c |c|c|c|}
    \hline
    \multirow{2}{*}{\textbf{\makecell{Loss \\Function}}} & \multicolumn{3}{c|}{\textbf{MalwareBazaar}} \\
    \cline{2-7} 
        & \textbf{ACC} & \textbf{F1} & \textbf{FNR} \\
    \hline 
        BCE  & 0.3640 & 0.5333 & 0.6360\\
        DRBCE & \textbf{0.4877} & \textbf{0.6552} & \textbf{0.5123}\\
    \hline
    \end{tabular}
    \label{tab:loss_function}
\end{table}

\subsection{Validation set}

It would be natural for a company to validate its model on the most recent samples that were collected, to ensure the model is robust against recent malware. We chose to assess the importance of the validation set by taking a random validation set from EMBER and a recent validation set composed of the very last $120,000$ samples of EMBER, and experimenting with both classical BCE loss and our DRBCE.

\begin{table}[!htb]
    \caption{Models performance comparison for different loss function and validation set combinations}
    \def\arraystretch{1.1}
    \centering 
    \begin{tabular}{ |c | c | c | c |}
    \hline
    \multirow{2}{*}{\textbf{\shortstack{ \\Validation set and loss \\function}}} & \multicolumn{3}{c|}{\textbf{BODMAS}} \\
    \cline{2-4} 
        & \textbf{ACC} & \textbf{F1-Score} & \textbf{FNR} \\
    \hline 
        BCE, random validation & 0.9247 & 0.9127 & 0.0769 \\
        BCE, recent validation & \textbf{0.9338 }& \textbf{0.9228} & 0.0696 \\ 
        \hdashline
        DRBCE, random validation & 0.9079 & 0.8995 & 0.0348 \\
        DRBCE, recent validation & 0.9128 & 0.9056 & 0.0193 \\
    \hline 
    \end{tabular}
\begin{tabular}{ |c | c | c | c | }
    \hline
    \multirow{2}{*}{\textbf{\shortstack{ \\Validation set and loss \\function}}} & \multicolumn{3}{c|}{\textbf{MalwareBazaar}}\\
    \cline{2-4} 
        & \textbf{ACC} & \textbf{F1} & \textbf{FNR}\\
    \hline 
        BCE, random validation &  0.3640 & 0.5333 &  0.6360\\
        BCE, recent validation & 0.3551 & 0.5238 & 0.6449\\ 
        \hdashline
        DRBCE, random validation & 0.4877 & 0.6552 & 0.5123\\
        DRBCE, recent validation & \textbf{0.4971} & \textbf{0.6634} &  \textbf{0.5029} \\
    \hline 
    \end{tabular}
    \label{tab:val_set}
\end{table}

As we can see in Table~\ref{tab:val_set}, the recent validation set does not give better long-term performance when using the BCE loss function, as it increases the accurcacy on BODMAS by 1.16\% and decreases the accuracy on the MalwareBazaar dataset by half a percent. When using our DRBCE loss function, the recent validation set seems much more profitable, as it increases the accuracy on BODMAS by 0.5\% and on the MalwareBazaar dataset by almost a percent. These results show that the use of a recent validation set is beneficial for the model., yielding better short-term predictions for the BCE loss function and overall better performance for the DRBCE loss function.

\subsection{Feature reduction}

We aim to evaluate whether feature reduction, using Permutation Feature Importance (described in \ref{sec:perm_feat_imp}), helps our model to achieve better performance by discarding non-necessary or even misleading features. We performed the PFI using the F1-Score metric as importance score, because accuracy alone can be misleading for unbalanced datasets. We chose to run the PFI on BODMAS, as it is the most recent dataset containing both malicious and benign files. In addition, performing the PFI on EMBER could lead to overfitting.
We applied this procedure on two different models, trained either using BCE loss or DRBCE loss. The DRBCE loss function is supposed to enable the network to use the provided features in a different way. Therefore, employing a distinct feature reduction seems a more rational choice.
After the process, we are left with $1,529$ features for the BCE model, and $1,478$ for the DRBCE model, reducing by almost half the number of features. We first compare the baseline model using BCE loss to the one trained on less features.

\begin{table}[!hb]
    \caption{Models performance comparison for model before and after feature reduction}
    \def\arraystretch{1.1}
    \centering 
    \begin{tabular}{ |c | c | c | c |}
    \hline
    \multirow{2}{*}{\textbf{\shortstack{ \\Feature reduction and loss \\function}}} & \multicolumn{3}{c|}{\textbf{BODMAS}}\\
    \cline{2-4} 
        & \textbf{ACC} & \textbf{F1-Score} & \textbf{FNR}\\
    \hline 
        BCE without f.r. &  0.9247 & 0.9127 & 0.0769 \\
        BCE with f.r. & \textbf{0.9510} & \textbf{0.9436} & 0.0389 \\
    \hdashline
        DRBCE without f.r. &  0.9079 & 0.8995 & \textbf{0.0348} \\
        DRBCE with f.r.  & 0.9262 & 0.9167 & 0.0468 \\
    \hline 
    \end{tabular}

       \begin{tabular}{ |c | c | c | c |}
    \hline
    \multirow{2}{*}{\textbf{\shortstack{ \\Feature reduction and loss \\function}}} &  \multicolumn{3}{c|}{\textbf{MalwareBazaar}}\\
    \cline{2-4} 
        & \textbf{ACC} & \textbf{F1-Score} & \textbf{FNR} \\
    \hline 
        BCE without f.r. & 0.3640 & 0.5333 & 0.6360\\
        BCE with f.r. & 0.3911 & 0.5616 & 0.6089\\
    \hdashline
        DRBCE without f.r.  & 0.4877 & 0.6552 & 0.5123\\
        DRBCE with f.r.  & \textbf{0.5031} & \textbf{0.6691} & \textbf{0.5083}\\
    \hline 
    \end{tabular}

    \label{tab:fi}
\end{table}

As we can see in Table \ref{tab:fi}, the feature reduction yields better long-term results for both BCE and DRBCE losses: a 2.71\% increase in accuracy for the BC.E and 1.54\% for the DRBCE loss function. We can also notice that the feature reduction significantly decreases the FNR for the BCE loss function by 3.8\%. These results indicate that the feature reduction improves our models predictions for both short and long-term detection.

\subsection{Final results}

Our final model consists of the baseline model, trained with the validation set made up of the latest EMBER data, with the Drift Resilient Binary Cross-Entropy loss and feature reduction using Permutation Feature Importance. Table \ref{tab:results_BODMAS} and \ref{tab:results_MWB} show the performance of models with gradual improvements.

\begin{table*}[!ht]
  \caption{Models performance comparison on BODMAS, for gradual improvements to the baseline}
\def\arraystretch{1.1}
\centering 
    \begin{tabular}{| c | c | c | c | c | c | c | c |}
      \hline
      \multicolumn{2}{|c|}{\textbf{Loss function}} & \multicolumn{2}{c|}{\textbf{Val. set}} & \multicolumn{2}{c|}{\textbf{Feature red.}} & \multicolumn{2}{c|}{\textbf{BODMAS}}\\
      \hline
      \textbf{BCE} & \textbf{DRBCE} & \textbf{Random} & \textbf{Recent} &\textbf{No F.R.} & \textbf{F.R.}& \textbf{ACC} & \textbf{F1-Score}\\
      \hline
       \checkmark & & \checkmark & & \checkmark & & 0.9247 & 0.9127 \\ 
       \hline 
        \checkmark & & & \checkmark & & \checkmark & \textbf{0.9557} & \textbf{0.9487} \\
      \hline
      & \checkmark & \checkmark & & \checkmark & &  0.9079 & 0.8995\\ 
        \hline
      & \checkmark & & \checkmark & \checkmark & & 0.9128 & 0.9056\\ 
      \hline 
       & \checkmark & & \checkmark & & \checkmark & 0.9258 & 0.9175\\ 
       \hline 
    \end{tabular}

  \label{tab:results_BODMAS}
\end{table*}

\begin{table*}[!ht]
\caption{Models performance comparison on MalwareBazaar, for gradual improvements to the baseline}
\def\arraystretch{1.1}
\centering 
    \begin{tabular}{| c | c | c | c | c | c | c | c |}
      \hline
      \multicolumn{2}{|c|}{\textbf{Loss function}} & \multicolumn{2}{c|}{\textbf{Val. set}} & \multicolumn{2}{c|}{\textbf{Feature red.}} & \multicolumn{2}{c|}{\textbf{MalwareBazaar}}\\
      \hline
      \textbf{BCE} & \textbf{DRBCE} & \textbf{Random} & \textbf{Recent} &\textbf{No F.R.} & \textbf{F.R.}& \textbf{ACC} & \textbf{F1-Score}\\
      \hline
       \checkmark & & \checkmark & & \checkmark & & 0.3640 & 0.5333 \\ 
       \hline 
        \checkmark & & & \checkmark & & \checkmark & 0.3679 & 0.5372 \\
      \hline
      & \checkmark & \checkmark & & \checkmark & & 0.4877 & 0.6552  \\ 
        \hline
      & \checkmark & & \checkmark & \checkmark & & 0.4971 & 0.6634 \\ 
      \hline 
       & \checkmark & & \checkmark & & \checkmark & \textbf{0.5155} & \textbf{0.6798}\\ 
       \hline 
    \end{tabular}
  
  \label{tab:results_MWB}
\end{table*}

\begin{figure*}[!ht]
    \centering 
    \subfloat[Baseline and BCE with f.r. and recent validation on BODMAS\label{fig:final_comp:a}]{
        \includegraphics[scale=0.25]{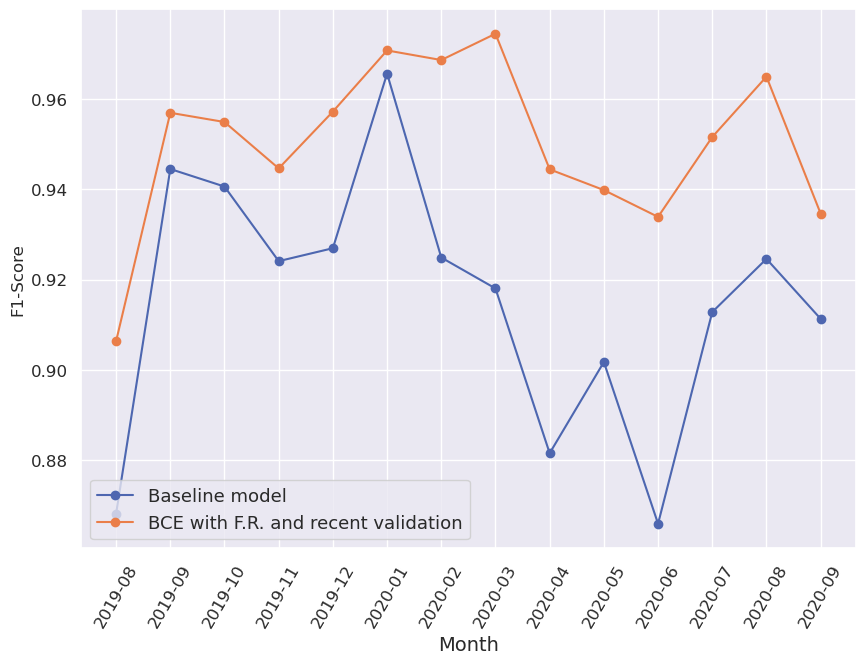}
    }
    \subfloat[Baseline and final model on MalwareBazaar\label{fig:final_comp:b}]{    
        \includegraphics[scale=0.25]{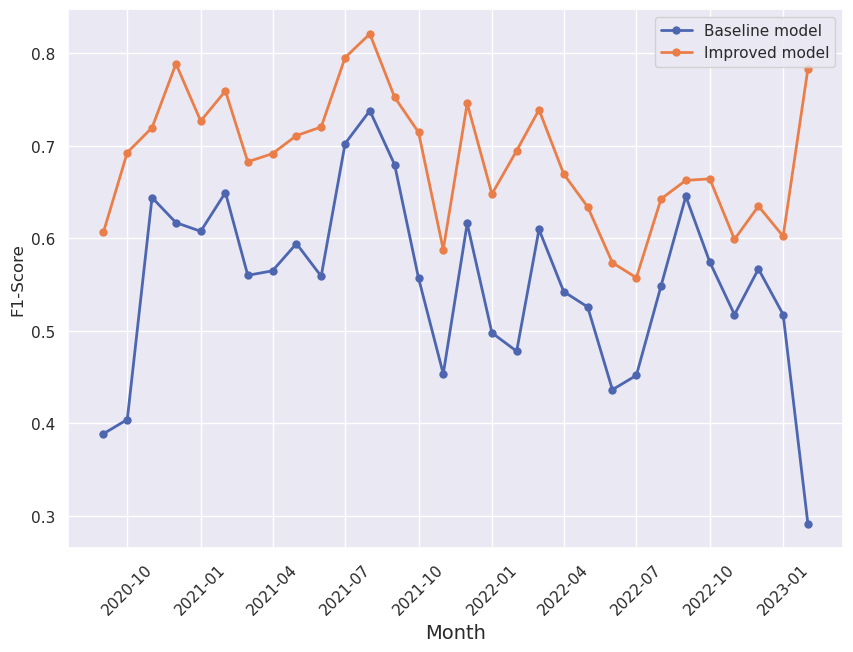}
    }
    \caption[F1-Score comparison between different models]{F1-Score comparison between different models}
    \label{fig:final_comp}
\end{figure*}

We can clearly see that our contributions have a great effect on drift reduction: our improved model achieves a much better result on recent files, with a 15.2\% increase in detection accuracy on the MalwareBazaar dataset. Each of our contributions seems to gradually improve the baseline model either on BODMAS or the MalwareBazaar dataset, showing that properly optimizing a model and using a custom drift resilient loss function leads to better long-term results, as shown in \ref{fig:final_comp:b}. It is also important to mention that the BCE loss function seems to benefit from the recent validation set and the feature reduction as it improves its F1-Score on short-term malware detection by 3.4\%, as shown in Figure \ref{fig:final_comp:a}. This implies that our custom loss function is more suited for long-term predictions while the BCE loss performs better for short-term malware detection.

\section{Conclusion and Discussions}
\label{sec:conclusion}

In this study, we propose three main improvements against concept and data drift in the malware detection field, and specifically for the EMBER set of features. Our first and main improvement is the Drift Resilient Binary Cross-Entropy, a loss function designed to help models to handle drift using spectral decoupling and penalization on false positive and false negative predictions. Our second improvement consists of using the most recent validation set available, and our third improvement is a feature reduction using the Permutation Feature Importance algorithm to discard non-beneficial features. All of our contributions aim to be easy to use and adaptable to the needs of each environment depending on the distribution of malicious and benign files in it. Our contributions are also model-agnostic and can be use with different algorithms and for other binary classification problems. 

Our contributions help our model handling the drift issue, and we want to highlight the fact that both loss functions (BCE and DRBCE) benefit from a recent validation set and feature reduction. The BCE loss function is more efficient with files chronologically close to the training dataset, while the DRBCE helps the model in performing better for recent files. Research has shown that using ensemble learning or aggregating models predictions can help when dealing with concept drift \cite{Lu2019,Hu2017}. Experiments should be carried out as part of further research to determine if combining multiple models trained using different loss functions could lead to greater robustness and better short and long-term detection performance.

A few points can be made about our model, methodology and datasets: firstly, the baseline model is a non-optimized Residual Neural Network that might not be fitted to this particular task. Our work is tailored to be model-agnostic, and the current baseline could be improved, enlarged and adapted to this specific classification task to yield better results. As we worked with regard to the model's performance exclusively, we directed our attention towards visualizing the model's predictions on a more recent dataset. Additional studies should be undertaken to assess the quantitative impact of our method on the drift.
We also point out the fact that the MalwareBazaar dataset does not contain recent benign files as they are complicated to collect, in particular because of copyright issues.

\clearpage

\bibliographystyle{unsrt}
\bibliography{biblio}

\end{document}